\documentclass{article}

\usepackage{microtype}
\usepackage{graphicx}
\usepackage{subfigure}
\usepackage{booktabs} 

\usepackage{hyperref}

\usepackage{amsfonts} 
\usepackage{amsmath}
\usepackage{mathabx}
\usepackage{bbm}
\usepackage[table]{xcolor}

\usepackage{listings}
\usepackage{enumitem}

\usepackage{xcolor}

\definecolor{codegreen}{rgb}{0,0.6,0}
\definecolor{codegray}{rgb}{0.5,0.5,0.5}
\definecolor{codepurple}{rgb}{0.58,0,0.82}
\definecolor{backcolour}{rgb}{0.96,0.96,0.96}

\lstdefinestyle{mystyle}{
    backgroundcolor=\color{backcolour},   
    commentstyle=\color{codegreen},
    keywordstyle=\color{magenta},
    numberstyle=\tiny\color{codegray},
    stringstyle=\color{codepurple},
    basicstyle=\ttfamily\footnotesize,
    breakatwhitespace=false,         
    breaklines=true,                 
    captionpos=b,                    
    keepspaces=true,                 
    numbers=left,                    
    numbersep=5pt,                  
    showspaces=false,                
    showstringspaces=false,
    showtabs=false,                  
    tabsize=2
}

\usepackage[accepted]{icml2024}

\icmltitlerunning{Deciphering RNA Secondary Structure Prediction: A Probabilistic K-Rook Matching Perspective}

\begin{document}

\twocolumn[
\icmltitle{Deciphering RNA Secondary Structure Prediction: A Probabilistic K-Rook Matching Perspective}



\icmlsetsymbol{equal}{*}

\begin{icmlauthorlist}
\icmlauthor{Cheng Tan}{zju,westlake,equal}
\icmlauthor{Zhangyang Gao}{westlake,zju,equal}
\icmlauthor{Hanqun Cao}{cuhk}
\icmlauthor{Xingran Chen}{umich}
\icmlauthor{Ge Wang}{westlake}
\icmlauthor{Lirong Wu}{westlake}
\icmlauthor{Jun Xia}{westlake}
\icmlauthor{Jiangbin Zheng}{westlake}
\icmlauthor{Stan Z. Li}{westlake}
\end{icmlauthorlist}

\icmlaffiliation{zju}{Zhejiang University}
\icmlaffiliation{westlake}{Westlake University}
\icmlaffiliation{cuhk}{The Chinese University of Hong Kong}
\icmlaffiliation{umich}{University of Michigan}

\icmlcorrespondingauthor{Stan Z. Li}{Stan.ZQ.Li@westlake.edu.cn}

\icmlkeywords{Machine Learning, ICML}

\vskip 0.3in
]



\printAffiliationsAndNotice{\icmlEqualContribution} 

\begin{abstract}
The secondary structure of ribonucleic acid (RNA) is more stable and accessible in the cell than its tertiary structure, making it essential for functional prediction. Although deep learning has shown promising results in this field, current methods suffer from poor generalization and high complexity. In this work, we reformulate the RNA secondary structure prediction as a K-Rook problem, thereby simplifying the prediction process into probabilistic matching within a finite solution space. Building on this innovative perspective, we introduce RFold, a simple yet effective method that learns to predict the most matching K-Rook solution from the given sequence. RFold employs a bi-dimensional optimization strategy that decomposes the probabilistic matching problem into row-wise and column-wise components to reduce the matching complexity, simplifying the solving process while guaranteeing the validity of the output. Extensive experiments demonstrate that RFold achieves competitive performance and about eight times faster inference efficiency than the state-of-the-art approaches. The code is available at \href{https://github.com/A4Bio/RFold}{github.com/A4Bio/RFold}.
\end{abstract}

\section{Introduction}
\begin{figure}[t]
    \centering
    \includegraphics[width=0.46\textwidth]{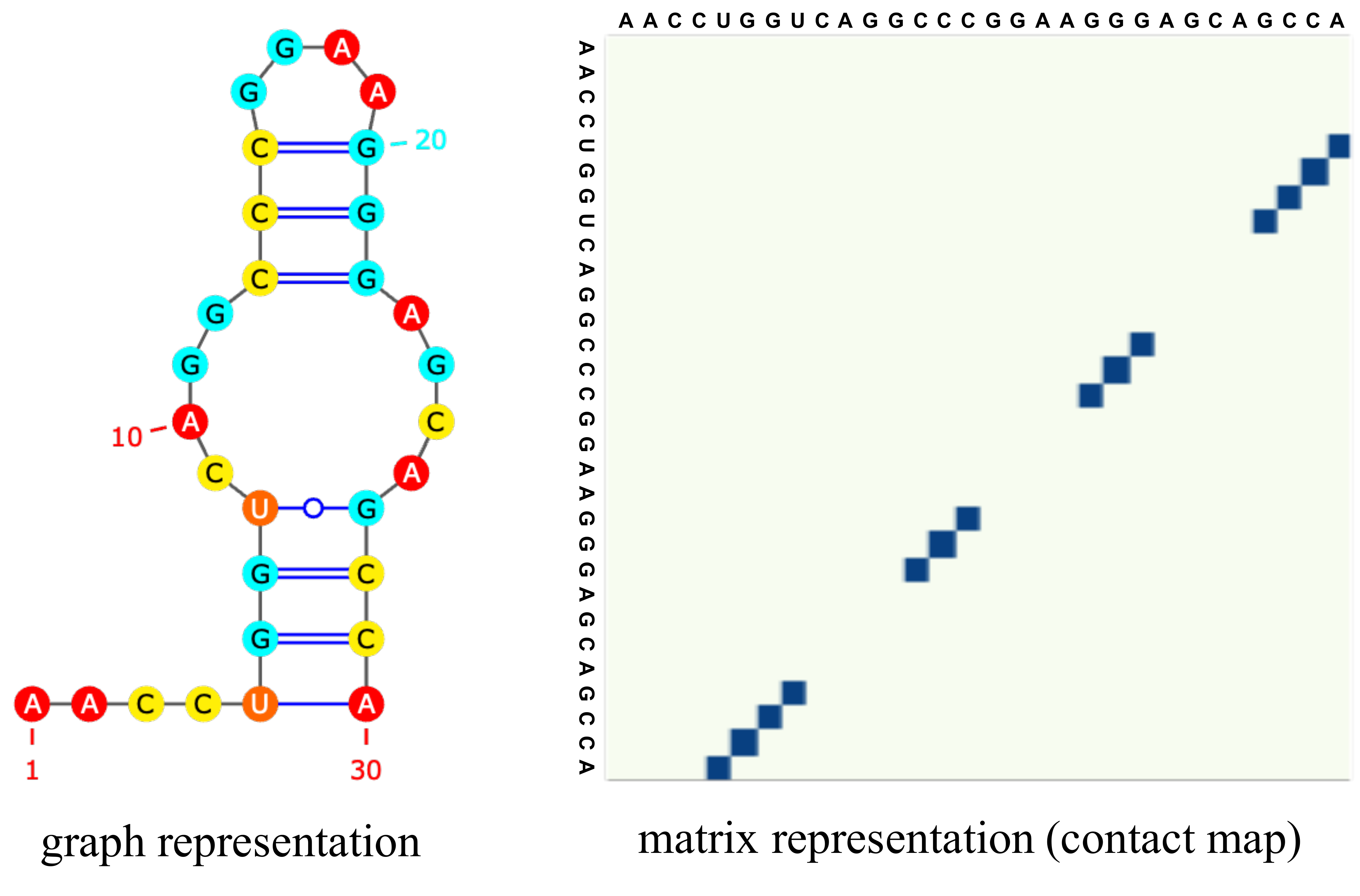}
    \vspace{-4mm}
    \caption{The graph and matrix representation of an RNA secondary structure example.}
    \vspace{-7mm}
    \label{fig:rna_sec_representations}
\end{figure}

The functions of RNA molecules are determined by their structure~\citep{sloma2016exact}. The secondary structure, which contains the nucleotide base pairing information, as shown in Figure~\ref{fig:rna_sec_representations}, is crucial for the correct functions of RNA molecules~\citep{fallmann2017recent}. Although experimental assays such as X-ray crystallography~\citep{cheong2004rapid}, nuclear magnetic resonance~\citep{furtig2003nmr}, and cryogenic electron microscopy~\citep{fica2017cryo} can be implemented to determine RNA secondary structure, they suffer from low throughput and expensive cost. 

\begin{figure}[ht]
    \centering
    \vspace{-2mm}
    \includegraphics[width=0.4\textwidth]{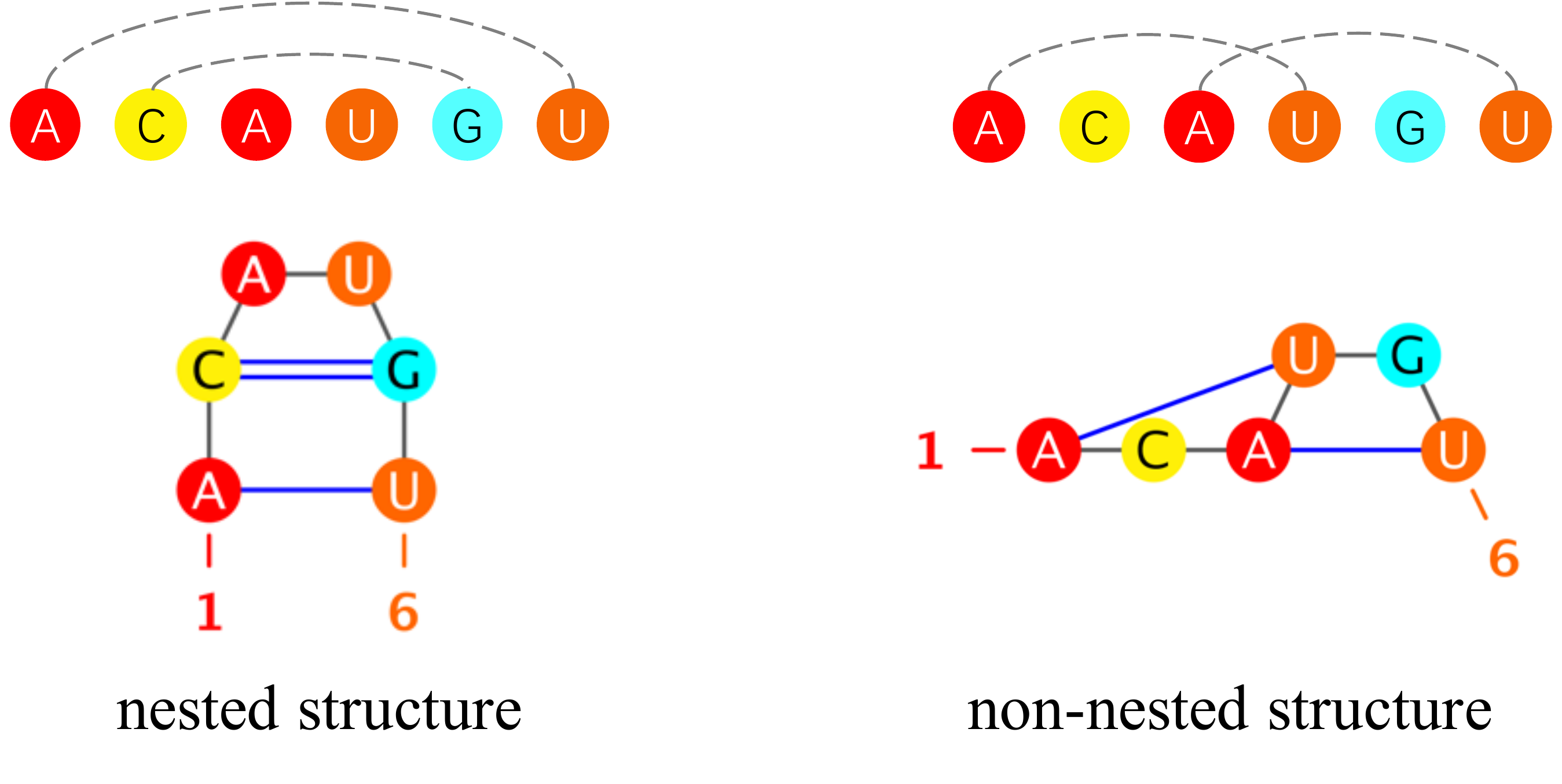}
    \vspace{-5mm}
    \caption{Examples of nested and non-nested secondary structures.}
    \vspace{-4mm}
    \label{fig:nested_non-nested}
\end{figure}

Computational RNA secondary structure prediction methods have been favored for their high efficiency in recent years~\cite{iorns2007utilizing}. Currently, mainstream methods can be broadly classified into two categories~\cite{rivas2013four,szikszai2022deep}: (i) comparative sequence analysis and (ii) single sequence folding algorithm. Comparative sequence analysis determines the secondary structure conserved among homologous sequences but the limited known RNA families hinder its development~\cite{gutell2002accuracy,griffiths2003rfam,gardner2009rfam,nawrocki2015rfam}. Researchers thus resort to single RNA sequence folding algorithms that do not need multiple sequence alignment information. A classical category of computational RNA folding algorithms is to use dynamic programming (DP) that assumes the secondary structure is a result of energy minimization~\cite{bellaousov2013rnastructure,nicholas2008unafold,lorenz2011viennarna,zuker2003mfold,mathews2006prediction,do2006contrafold}. However, energy-based approaches usually require the base pairs have a nested structure while ignoring some valid yet biologically essential structures such as pseudoknots, i.e., non-nested base pairs~\cite{e2efold,seetin2012rna,xu2015physics}, as shown in Figure~\ref{fig:nested_non-nested}. Since predicting secondary structures with pseudoknots under the energy minimization framework has shown to be hard and NP-complete~\cite{wang2011dynamic,ufold}, deep learning techniques are introduced as an alternative. 

Attempts to overcome the limitations of energy-based methods have motivated deep learning methods that predict RNA secondary structures in the absence of DP. SPOT-RNA~\cite{spot-rna} is a seminal work that ensembles ResNet~\cite{he2016deep} and LSTM~\cite{hochreiter1997long} and applies transfer learning to identify molecular recognition features. SPOT-RNA does not constrain the output space into valid RNA secondary structures, which degrades its generalization ability on new datasets~\cite{rtfold}. E2Efold~\cite{e2efold} employs an unrolled algorithm for constrained programming that post-processes the network output to satisfy the constraints. E2Efold introduces a convex relaxation to make the constrained optimization tractable, leading to possible structural constraint violations and poor generalization ability~\cite{sato2021rna,ufold,franke2023scalable,franke2022probabilistic}. RTfold~\cite{rtfold} utilizes the Fenchel-Young loss~\cite{berthet2020learning} to enable differentiable discrete optimization with perturbations, but the approximation cannot guarantee the satisfaction of constraints. Developing an appropriate optimization that enforces the output to be valid becomes a crucial concern.

Since deep learning-based approaches cannot directly output valid RNA secondary structures, existing approaches usually formulate the problem into a constrained optimization problem and optimize the output of the model to fulfill specific constraints as closely as possible. However, these methods typically necessitate iterative optimization, leading to reduced efficiency. Moreover, the extensive optimization space involved does not ensure the complete satisfaction of these constraints. In this study, we introduce a novel perspective for predicting RNA secondary structures by reframing the challenge as a K-Rook problem. Recognizing the alignment between the solution spaces of the K-Rook problem and RNA secondary structure prediction, our objective is to identify the most compatible K-Rook solution for each RNA sequence. This is achieved by training the deep learning model on prior data to learn matching patterns. 

Considering the high complexity of the solution space in the symmetric K-Rook problem, we introduced RFold, an innovative approach. This method utilizes a bi-dimensional optimization strategy, effectively decomposing the problem into separate row-wise and column-wise components. This decomposition significantly reduces the matching complexity, thereby simplifying the solving process while guaranteeing the validity of the output. We conduct extensive experiments to compare RFold with state-of-the-art methods on several benchmark datasets and show the superior performance of our proposed method. Moreover, RFold has faster inference efficiency than those methods due to its simplicity.

\vspace{-3mm}
\section{Related work}
\vspace{-1mm}

\paragraph{Comparative Sequence Analysis}
Comparative sequence analysis determines base pairs conserved among homologous sequences~\citep{gardner2004comprehensive,knudsen2003pfold,gorodkin2001discovering}. ILM~\citep{ruan2004iterated} combines thermodynamic and mutual information content scores. Sankoff~\citep{hofacker2004alignment} merges the sequence alignment and maximal-pairing folding methods~\citep{nussinov1978algorithms}. Dynalign~\citep{mathews2002dynalign} and Carnac~\citep{touzet2004carnac} are the subsequent variants of Sankoff algorithms. RNA forester~\citep{hochsmann2003local} introduces a tree alignment model for global and local alignments. However, the limited number of known RNA families~\citep{nawrocki2015rfam} impedes the development.

\vspace{-4mm}
\paragraph{Energy-based Folding Algorithms}
When the structures consist only of nested base pairing, dynamic programming can predict the structure by minimizing energy. Early works include Vienna RNAfold~\citep{lorenz2011viennarna}, Mfold~\citep{zuker2003mfold}, RNAstructure~\citep{mathews2006prediction}, and CONTRAfold~\citep{do2006contrafold}. Faster implementations that speed up dynamic programming have been proposed, such as Vienna RNAplfold~\citep{bernhart2006local}, LocalFold~\citep{lange2012global}, and LinearFold~\citep{huang2019linearfold}. However, they cannot accurately predict structures with pseudoknots, as predicting the lowest free energy structures with pseudoknots is NP-complete~\citep{lyngso2000rna}, making it difficult to improve performance.

\vspace{-4mm}
\paragraph{Learning-based Folding Algorithms} Deep learning methods have inspired approaches in bioengineering applications~\cite{wu2024psc,wu2024mape,Lin2022DiffBPGD,Lin2023FunctionalGroupBasedDF,tan2024cross,tan2023hierarchical}.
SPOT-RNA~\citep{spot-rna} is a seminal work that employs deep learning for RNA secondary structure prediction. SPOT-RNA2~\citep{singh2021improved} improves its predecessor by using evolution-derived sequence profiles and mutational coupling. Inspired by Raptor-X~\citep{wang2017accurate} and SPOT-Contact~\citep{hanson2018accurate}, SPOT-RNA uses ResNet and bidirectional LSTM with a sigmoid function. MXfold~\citep{akiyama2018max} combines support vector machines and thermodynamic models. CDPfold~\citep{zhang2019new}, DMFold~\citep{wang2019dmfold}, and MXFold2~\citep{sato2021rna} integrate deep learning techniques with energy-based methods. E2Efold~\citep{e2efold} constrains the output to be valid by learning unrolled algorithms. However, its relaxation for making the optimization tractable may violate the constraints. UFold~\citep{ufold} introduces U-Net model to improve performance.

\section{Background}

\subsection{Preliminaries}

The primary structure of RNA is a sequence of nucleotide bases A, U, C, and G, arranged in order and represented as $\boldsymbol{X} = (x_1, ..., x_L)$, where each $x_i$ denotes one of these bases. The secondary structure is the set of base pairings within the sequence, modeled as a sparse matrix $\boldsymbol{M} \in \{0,1\}^{L \times L}$, where $\boldsymbol{M}_{ij} = 1$ indicates a bond between bases $i$ and $j$. The key challenges include (i) designing a model, characterized by parameters $\Theta$, that captures the complex transformations from the sequence $\boldsymbol{X}$ to the pairing matrix $\boldsymbol{M}$ and (ii) correctly identifying the sparsity of the secondary structure, which is determined by the nature of RNA. Thus, the transformation $\mathcal{F}_\Theta: \boldsymbol{X} \mapsto \boldsymbol{M}$ is usually decomposed into two stages for capturing the sequence-to-structure relationship and optimizing the sparsity of the target matrix:
\begin{equation}
    \mathcal{F}_\Theta := \mathcal{G}_{\theta_g} \circ \mathcal{H}_{\theta_h},
\end{equation}
where $\mathcal{H}_{\theta_h}: \boldsymbol{X} \mapsto \boldsymbol{H}$ represents the initial processing step, transforming the RNA sequence into an intermediate, unconstrained representation $\boldsymbol{H} \in \mathbb{R}^{L \times L}$. Subsequently, $\mathcal{G}_{\theta_g}: \boldsymbol{H} \mapsto \boldsymbol{M}$ parameterizes the optimization stage for the intermediate distribution into the final sparse matrix $\boldsymbol{M}$. 

\subsection{Constrained Optimization-based Approaches}
The core problem of secondary structure prediction lies in sparsity identification. Numerous studies regard this task as a constrained optimization problem, seeking the optimal refinement mappings by gradient descent. Besides, keeping the hard constraints on RNA secondary structures is also essential, which ensures valid biological functions~\cite{steeg1993neural}. These constraints can be formally described as:
\begin{itemize}[leftmargin=*,itemsep=0pt,topsep=0pt]
\label{sec:constraints}
\item (a) Only three types of nucleotide combinations can form base pairs: $\mathcal{B} := \{\mathrm{AU}, \mathrm{UA}\} \cup \{\mathrm{GC}, \mathrm{CG}\} \cup \{\mathrm{GU}, \mathrm{UG}\}$. For any base pair $x_i x_j$ where $x_i x_j \notin \mathcal{B}$, $\boldsymbol{M}_{ij} = 0$. \label{constraint_a}
\item (b) No sharp loop within three bases. For any adjacent bases within a distance of three nucleotides, they cannot form pairs with each other. For all $|i-j| < 4, \boldsymbol{M}_{ij} = 0$. \label{constraint_b}
\item (c) There can be at most one pair for each base. For all $i$ and $j$, $\sum_{j=1}^L \boldsymbol{M}_{ij} \leq 1, \sum_{i=1}^L \boldsymbol{M}_{ij} \leq 1$. \label{constraint_c}
\end{itemize}
The search for valid secondary structures is thus a quest for \textit{symmetric} sparse matrices $\in \{0, 1\}^{L \times L}$ that adhere to the constraints above. The first two constraints can be satisfied by defining a constraint matrix $\boldsymbol{\widebar{M}}$ as: $\boldsymbol{\widebar{M}}_{ij}:=1$ if $x_i x_j \in \mathcal{B}$ and $|i-j| \geq 4$, and $\boldsymbol{\widebar{M}}_{ij}:=0$ otherwise. Addressing the third constraint is critical as it necessitates employing sparse optimization techniques. Therefore, our primary objective is to devise an effective sparse optimization strategy. This strategy is based on the symmetric inherent distribution $\boldsymbol{H}$ and $\boldsymbol{M}$, which support constraints (a) and (b), and additionally addresses constraint (c).

\paragraph{SPOT-RNA} subtly enforces the principles of sparsity. It streamlines the pathway from the raw neural network output $\boldsymbol{H}$ by harnessing the $\mathrm{Sigmoid}$ function to distill a sparse pattern. The transformation applies a threshold to yield a binary sparse matrix. This process can be represented as:
\begin{equation}
\mathcal{G}(\boldsymbol{H}) = \mathbbm{1}_{[\mathrm{Sigmoid}(\boldsymbol{H}) > s]}.
\end{equation}
In this approach, a fixed threshold $s$ of 0.5 is applied, typical for inducing sparsity. It omits complex constraints or extra parameters $\theta_g$, simply using this cutoff to achieve sparse structure representations.

\paragraph{E2Efold} introduces a non-linear transformation to the intermediate value $\boldsymbol{\widehat{M}} \in \mathbb{R}^{L \times L}$ and an additional regularization term $\|\boldsymbol{\widehat{M}}\|_1$. 
\begin{equation}
    \frac{1}{2} \left\langle \boldsymbol{H}-s, \mathcal{T}(\boldsymbol{\widehat{M}})\right\rangle - \rho \|\boldsymbol{\widehat{M}}\|_1,
    \label{eq:e2efold_ori}
\end{equation}
where $\mathcal{T}(\boldsymbol{\widehat{M}}) = \frac{1}{2}(\boldsymbol{\widehat{M}} \odot \boldsymbol{\widehat{M}} + (\boldsymbol{\widehat{M}} \odot \boldsymbol{\widehat{M}})^T) \odot \boldsymbol{\widebar{M}}$ ensures symmetry and adherence to RNA base-pairing constraints (a) and (b), $s$ is the log-ratio bias term set to $\log(9.0)$, and the $\ell_1$ penalty $\rho \|\boldsymbol{\widehat{M}}\|_1$ promotes sparsity. To fulfill constraint (c), the objective is combined with conditions $\mathcal{T}(\boldsymbol{\widehat{M}}) \mathbbm{1} \leq \mathbbm{1}$. Denote $\boldsymbol{\lambda} \in \mathbb{R}^{L}_+$ as the Lagrange multiplier, the formulation for the sparse optimization is expressed as:
\begin{equation}
\begin{aligned}
    \min_{\boldsymbol{\lambda} \geq \boldsymbol{0}} \max_{\boldsymbol{\widehat{M}}} \; &\frac{1}{2} \left\langle \boldsymbol{H}-s, \mathcal{T}(\boldsymbol{\widehat{M}})\right\rangle - \rho \|\boldsymbol{\widehat{M}}\|_1\\ &- \left\langle \boldsymbol{\lambda}, \mathrm{ReLU}(\mathcal{T}(\boldsymbol{\widehat{M}}) \mathbbm{1} - \mathbbm{1}) \right\rangle,
\end{aligned}
\end{equation}
In the training stage, the optimization objective is the output of score function $\mathcal{S}$ dependent on $\boldsymbol{\widehat{M}}$ and $\boldsymbol{H}$. It can be regarded as an optimization function $\mathcal{G}$ parameterized by $\theta_g$:
\begin{equation}
    \mathcal{G}_{\theta_g}(\boldsymbol{H}) = \mathcal{T}({\arg\max}_{\boldsymbol{\widehat{M}} \in \mathbb{R}^{L \times L}} \mathcal{S}(\boldsymbol{\widehat{M}}, \boldsymbol{H})).
\end{equation}
Although the complicated design to the constraints is explicitly formulated, the iterative updates may fall into sub-optimal or invalid solutions. Besides, it requires additional parameters $\theta_g$, making the model training complicated.  

\paragraph{RTfold} introduces a differentiable function that incorporates an additional Gaussian perturbation $\boldsymbol{W}$. The objective function is expressed as:
\begin{equation}
    \min_{\boldsymbol{\widehat{M}}} \frac{1}{N} \sum_{i=1}^{N} \mathcal{T}\left(\boldsymbol{H} + \epsilon \boldsymbol{W}^{(i)} \right) -\boldsymbol{\widehat{M}}
\end{equation}
where $\mathcal{T}$ denotes the non-linear transformation to constrain the initial output $\boldsymbol{H}$, and $N$ is the number of random samples. The random perturbation $\boldsymbol{W}^{(i)}$ adjusts the distribution by leveraging the gradient during the optimization process. While RTFold designs an efficient differential objective function, the constraints imposed by the non-linear transformation on a noisy hidden distribution may lead to biologically implausible structures.

\section{RFold}

\subsection{Probabilistic K-Rook Matching}

\begin{figure}[ht]
    \centering
    \includegraphics[width=0.48\textwidth]{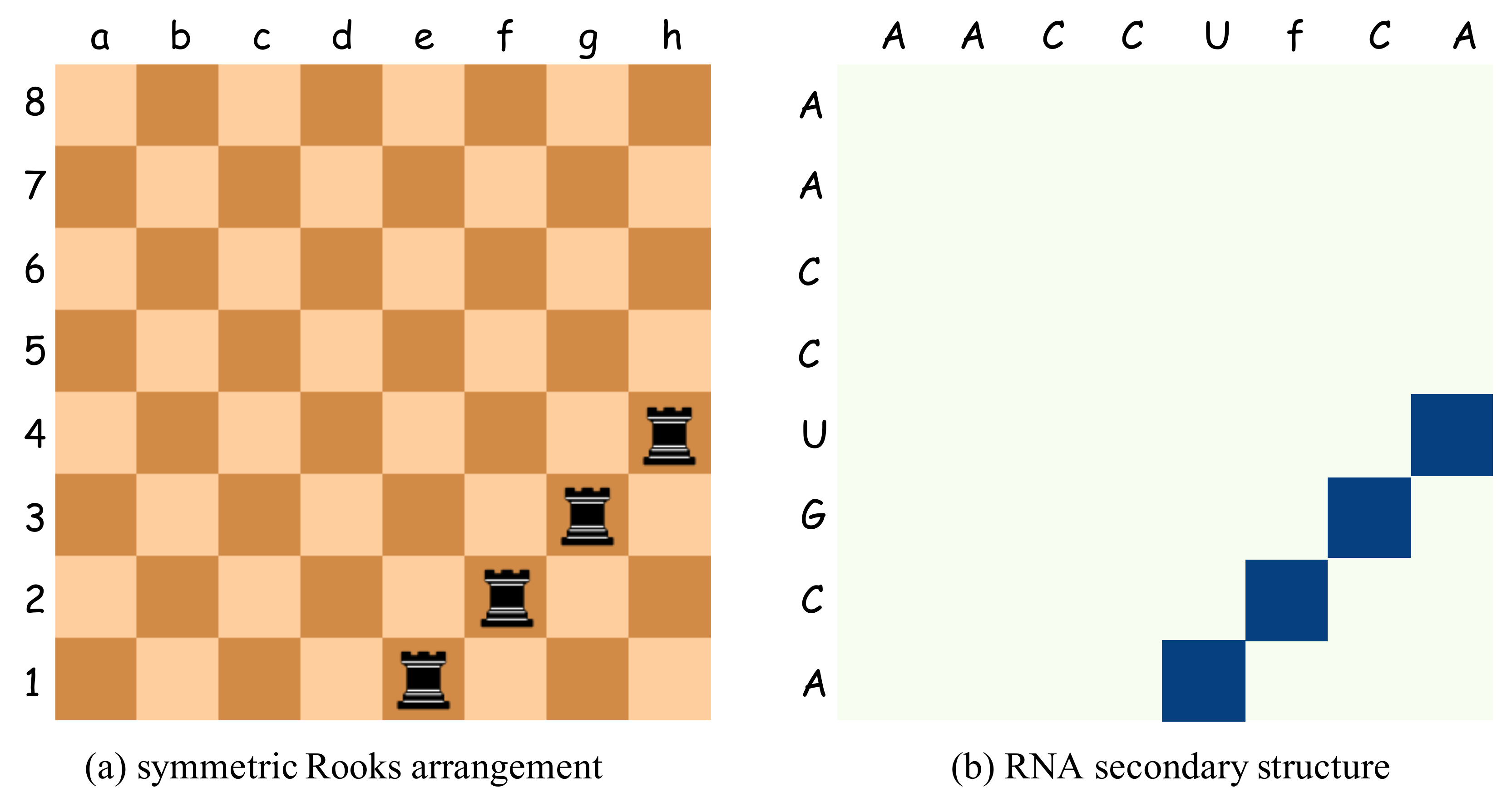}
    \vspace{-6mm}
    \caption{The analogy between the symmetric K-Rook arrangement and the RNA secondary structure prediction.}
    \label{fig:rooks}
\end{figure}

The symmetric K-Rook arrangement~\cite{riordan2014introduction,elkies2011chess} is a classic combinatorial problem involving the placement of $K (K \leq L)$ non-attacking Rooks on an $L\times L$ chessboard, where the goal is to arrange the Rooks such that they form a symmetric pattern. The term 'non-attacking' means that no two Rooks are positioned in the same row or column. An interesting parallel can be drawn between this combinatorial scenario and the domain of RNA secondary structure prediction, as illustrated in Figure~\ref{fig:rooks}. This analogy stems from the conceptual similarity in the arrangement patterns required in both cases. The RNA sequence can be regarded as a chessboard of size $L$ and the base pairs are the Rooks. The core problem is to determine an optimal arrangement of these base pairs.

Given a finite solution space $\mathcal{M}$ defined by the symmetric K-Rook arrangement, we reformulate our objective as a probabilistic matching problem. The goal is to find the most matching solution $\boldsymbol{M}^* \in \mathcal{M}$ for the given sequence $\boldsymbol{X}$. The optimal solution $\boldsymbol{M}^*$ is defined as:
\begin{equation}
\begin{aligned}
    \boldsymbol{M}^* &= \arg\max_{\boldsymbol{M} \in \mathcal{M}} \mathcal{P}(\boldsymbol{M} | \boldsymbol{X}).
\end{aligned}
\end{equation}

According to Bayes' theorem, the posterior probability can be represented as $\mathcal{P}(\boldsymbol{M} | \boldsymbol{X}) = \frac{\mathcal{P}(\boldsymbol{X} | \boldsymbol{M}) \mathcal{P}(\boldsymbol{M})}{\mathcal{P}(\boldsymbol{X})}$. Since the denominator $P(\boldsymbol{X})$ is constant for all $\boldsymbol{M}$, and assuming that the solution space is finite and each solution within it is equally likely, we can adopt a uniform prior $\mathcal{P}(\boldsymbol{M})$ in this context. Therefore, maximizing the posterior probability is equivalent to maximizing the likelihood $\mathcal{P}(\boldsymbol{X}|\boldsymbol{M})$. This leads to the following equation:
\begin{equation}
    \boldsymbol{M}^* = \arg\max_{\boldsymbol{M} \in \mathcal{M}} \mathcal{P}(\boldsymbol{X}|\boldsymbol{M}).
\end{equation}
Therefore, our primary task becomes computing the likelihood $\mathcal{P}(\boldsymbol{X}|\boldsymbol{M})$ for the given sequence $\boldsymbol{X}$ under each possible solution $\boldsymbol{M}$.

\subsection{Bi-dimensional Optimization}

Computing the likelihood $\mathcal{P}(\boldsymbol{X} | \boldsymbol{M})$ directly poses significant challenges. To address this, we propose a bi-dimensional optimization strategy that simplifies the problem by decomposing it into row-wise and column-wise components. This approach is mathematically represented as:
\begin{equation}
\mathcal{P}(\boldsymbol{X}|\boldsymbol{M}) = \mathcal{P}(\boldsymbol{X}|\boldsymbol{R}) \mathcal{P}(\boldsymbol{X}|\boldsymbol{C}),
\end{equation}
where $\boldsymbol{M}$ is the product of the row-wise component $\boldsymbol{R} \in \mathbb{R}^{L\times L}$ and the column-wise component $\boldsymbol{C} \in \mathbb{R}^{L\times L}$, i.e., $\boldsymbol{M} = \boldsymbol{R} \odot \boldsymbol{C}$. Each component represents the optimal solution for the row-wise and column-wise matching problems, respectively. Importantly, the row-wise and column-wise components are independent, and the comprehensive solution for the entire problem is derived from the product of the optimal solutions for these two sub-problems.

Applying Bayes' theorem, for the row-wise component, we have $\mathcal{P}(\boldsymbol{R} | \boldsymbol{X}) = \frac{\mathcal{P}(\boldsymbol{X} | \boldsymbol{R}) \mathcal{P}(\boldsymbol{R})}{\mathcal{P}(\boldsymbol{X})}$. Given that the solution space of $\boldsymbol{R}$ is both finite and valid, we can regard it as a uniform distribution. The analysis for the column-wise component, $\mathcal{P}(\boldsymbol{C} | \boldsymbol{X})$, follows a similar approach. Therefore, the optimal solution $\boldsymbol{M}^*$ can be represented as:
\begin{equation}
\begin{aligned}
\boldsymbol{M}^* &= \arg\max_{\boldsymbol{R}, \boldsymbol{C}} \mathcal{P}(\boldsymbol{R}|\boldsymbol{X}) \mathcal{P}(\boldsymbol{C}|\boldsymbol{X}) \\
&= \arg\max_{\boldsymbol{R}} \mathcal{P}(\boldsymbol{R}|\boldsymbol{X}) \arg\max_{\boldsymbol{C}}\mathcal{P}(\boldsymbol{C}|\boldsymbol{X})
\end{aligned}
\end{equation}
The next phase involves establishing proxies for $\mathcal{P}(\boldsymbol{R}|\boldsymbol{X})$ and $\mathcal{P}(\boldsymbol{C}|\boldsymbol{X})$. To this end, we introduce the basic symmetric hidden distribution, $\boldsymbol{\widehat{H}} = (\boldsymbol{H} \odot \boldsymbol{H}^T) \odot \boldsymbol{\widebar{M}}$. The row-wise and column-wise components are then derived by applying Softmax functions to $\boldsymbol{\widehat{H}}$, resulting in their respective probability distributions:
\begin{equation}
\begin{aligned}
\mathcal{R}(\boldsymbol{\widehat{H}}) = \frac{\exp(\boldsymbol{\widehat{H}}_{ij})}{\sum_{k=1}^L\exp(\boldsymbol{\widehat{H}}_{ik})}, \mathcal{C}(\boldsymbol{\widehat{H}}) = \frac{\exp(\boldsymbol{\widehat{H}}_{ij})}{\sum_{k=1}^L\exp(\boldsymbol{\widehat{H}}_{kj})}.
\end{aligned}
\end{equation}
The final output is the element-wise product of the row-wise component $\mathcal{R}(\boldsymbol{\widehat{H}})$ and the column-wise component $\mathcal{C}(\boldsymbol{\widehat{H}})$. This operation integrates the individual insights from both dimensions, leading to the optimized matrix $\boldsymbol{M}^*$:
\begin{equation}
    \boldsymbol{M}^* = \arg\max \mathcal{R}(\boldsymbol{\widehat{H}}) \odot \arg\max \mathcal{C}(\boldsymbol{\widehat{H}}).
\end{equation}

\begin{figure}[ht]
    \centering
    \vspace{-2mm}
    \includegraphics[width=0.48\textwidth]{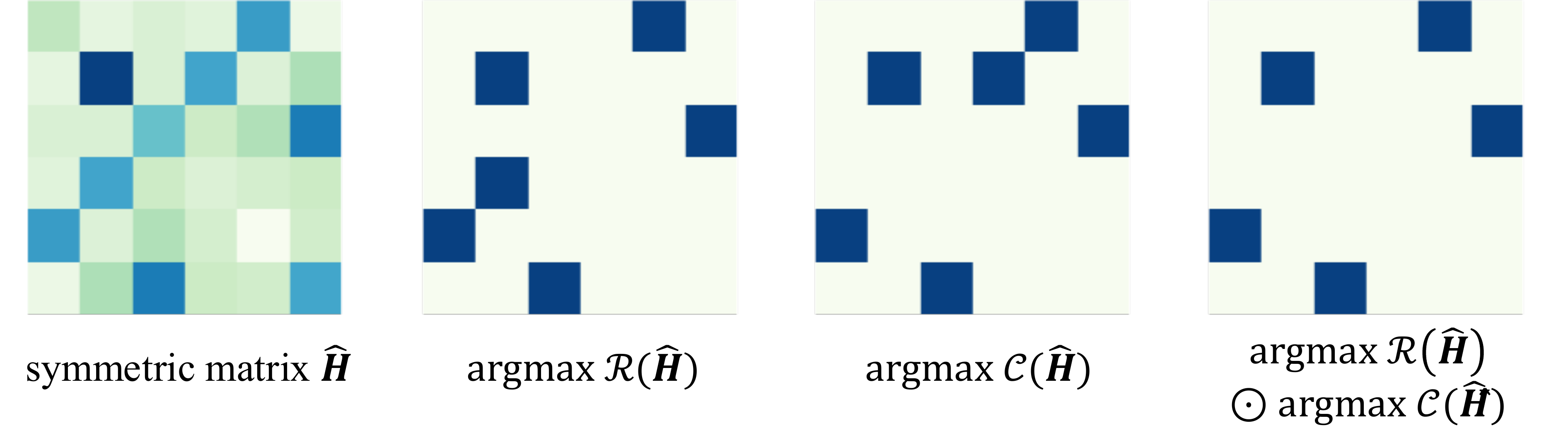}
    \vspace{-4mm}
    \caption{The visualization of $\arg\max\mathcal{R}(\boldsymbol{\widehat{H}}) \odot \arg\max\mathcal{C}(\boldsymbol{\widehat{H}})$.}
    \label{fig:row_col_argmax}
\end{figure}

As illustrated in Figure~\ref{fig:row_col_argmax}, we consider a random symmetric $6\times 6$ matrix as an example. For simplicity, we disregard the constraints (a-b) from $\boldsymbol{\widebar{M}}$. This example demonstrates the outputs of $\mathcal{R}(\cdot)$, $\mathcal{C}(\cdot)$, and their element-wise product $\mathcal{R}(\cdot) \odot \mathcal{C}(\cdot)$. The row-wise and column-wise components jointly select the value that has the maximum in both its row and column while keeping the output matrix symmetric.

Given the definition of $\boldsymbol{\widehat{H}} = (\boldsymbol{H} \odot \boldsymbol{H}^T) \odot \boldsymbol{\widebar{M}}$, it is evident that $\boldsymbol{\widehat{H}}$ inherently forms a symmetric and non-negative matrix. Regarding optimization, the operation $\mathcal{R}(\boldsymbol{\widehat{H}}) \odot \mathcal{C}(\boldsymbol{\widehat{H}})$ can be equivalently simplified to optimizing $\frac{1}{2}(\mathcal{R}(\boldsymbol{\widehat{H}}) + \mathcal{C}(\boldsymbol{\widehat{H}}))$. This is because both approaches fundamentally aim to maximize the congruence between the row-wise and column-wise components of $\boldsymbol{\widehat{H}}$. The underlying reason for this equivalence is that both optimizing the Hadamard product and the arithmetic mean of $\mathcal{R}(\boldsymbol{\widehat{H}})$ and $\mathcal{C}(\boldsymbol{\widehat{H}})$ focus on reinforcing the alignment and coherence across the various dimensions of the matrix.

Moreover, examining the gradients of these operations sheds light on their computational efficiencies. The gradient of $\mathcal{R}(\boldsymbol{\widehat{H}}) \odot \mathcal{C}(\boldsymbol{\widehat{H}})$ entails a blend of partial derivatives interconnected via element-wise multiplication. It can be formally expressed as follows:
\begin{equation}
\begin{aligned}
    &\frac{\partial(\mathcal{R}(\boldsymbol{\widehat{H}}) \odot \mathcal{C}(\boldsymbol{\widehat{H}}))_{ij}}{\partial \boldsymbol{\widehat{H}}_{ij}} \\ = &\mathcal{C}(\boldsymbol{\widehat{H}})_{ij} \cdot \frac{\partial \mathcal{R}(\boldsymbol{\widehat{H}})_{ij}}{\partial \boldsymbol{\widehat{H}}_{ij}} + \mathcal{R}(\boldsymbol{\widehat{H}})_{ij} \cdot \frac{\partial \mathcal{C}(\boldsymbol{\widehat{H}})_{ij}}{\partial \boldsymbol{\widehat{H}}_{ij}}.
    \\
\end{aligned}
\end{equation}
In contrast, the gradient of $\frac{1}{2}(\mathcal{R}(\boldsymbol{\widehat{H}}) + \mathcal{C}(\boldsymbol{\widehat{H}}))$ is characterized by a straightforward sum of partial derivatives:
\begin{equation}
    \begin{aligned}
        \frac{\partial (\mathcal{R}(\boldsymbol{\widehat{H}}) + \mathcal{C}(\boldsymbol{\widehat{H}}))_{ij}}{\partial \boldsymbol{\widehat{H}}_{ij}} = \frac{\partial \mathcal{R}(\boldsymbol{\widehat{H}})_{ij}}{\partial \boldsymbol{\widehat{H}}_{ij}} + \frac{\partial \mathcal{C}(\boldsymbol{\widehat{H}})_{ij}}{\partial \boldsymbol{\widehat{H}}_{ij}}.
    \end{aligned}
\end{equation}
Element-wise addition, as used in the latter, tends to be numerically more stable and less susceptible to issues like floating-point precision errors, which are more common in element-wise multiplication operations. This stability is particularly beneficial when dealing with large-scale matrices or when the gradients involve extreme values, where numerical instability can pose significant challenges.

The proposed simplification not only maintains the mathematical integrity of the optimization problem but also provides computational advantages, making it a desirable strategy in practical scenarios involving large and intricate datasets. Consequently, we define the overall loss function as the mean square error (MSE) between the averaged row-wise and column-wise components of $\boldsymbol{\widehat{H}}$ and the ground truth secondary structure $\boldsymbol{M}$:
\begin{equation}
    \mathcal{L}(\boldsymbol{M}^*, \boldsymbol{M}) = \frac{1}{L^2} \Big\|\frac{1}{2}(\mathcal{R}(\boldsymbol{\widehat{H}}) + \mathcal{C}(\boldsymbol{\widehat{H}})) - \boldsymbol{M}\Big\|^2.
\end{equation}

\subsection{Practical Implementation}

\begin{figure}[h]
    \centering
    \includegraphics[width=0.48\textwidth]{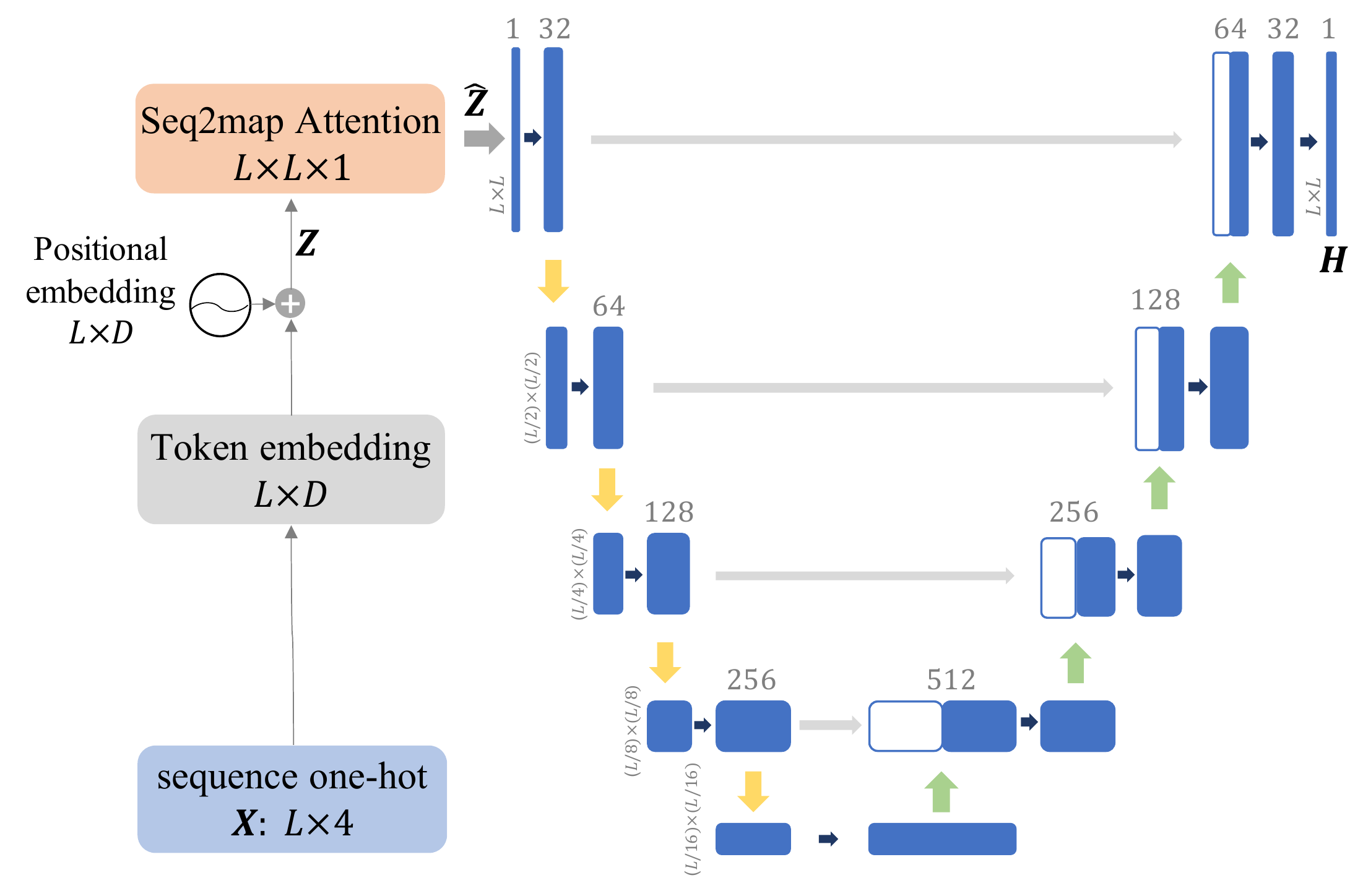}
    \vspace{-2mm}
    \caption{The overview model architecture of RFold.}
    \label{fig:overview}
\end{figure}

We identify the problem of predicting $\boldsymbol{H} \in \mathbb{R}^{L \times L}$ from the given sequence attention map $\boldsymbol{\widehat{Z}} \in \mathbb{R}^{L \times L}$ as an image-to-image segmentation problem and apply the U-Net model to extract pair-wise information, as shown in Figure~\ref{fig:overview}.

To automatically learn informative representations from sequences, we propose a Seq2map attention module. Given a sequence in one-hot form $\boldsymbol{X} \in \mathbb{R}^{L\times 4}$, we first obtain the sum of the token embedding and positional embedding as the input of the Seq2map attention. We denote the input as $\boldsymbol{Z}\in \mathbb{R}^{L \times D}$ for convenience, where $D$ is the hidden layer size of the token and positional embeddings. 

Motivated by the recent progress in attention mechanisms~\citep{vaswani2017attention,choromanski2020rethinking,katharopoulos2020transformers,hua2022transformer}, we aim to develop a highly effective sequence-to-map transformation based on pair-wise attention. We obtain the query $\boldsymbol{Q} \in \mathbb{R}^{L \times D}$ and key $\boldsymbol{K} \in \mathbb{R}^{L \times D}$ by
applying per-dim scalars and offsets to $\boldsymbol{Z}$:
\begin{equation}
\begin{aligned}
    \boldsymbol{Q} &= \gamma_Q \boldsymbol{Z} + \beta_Q, \\
    \boldsymbol{K} &= \gamma_K \boldsymbol{Z} + \beta_K,
\end{aligned}
\end{equation}
where $\gamma_Q,  \gamma_K, \beta_Q, \beta_K \in \mathbb{R}^{L \times D}$ are learnable parameters.

Then, the pair-wise attention map is obtained by:
\begin{equation}
    \boldsymbol{\widebar{Z}} = \mathrm{ReLU}^2(\boldsymbol{Q} \boldsymbol{K}^T / L),
\end{equation}
where $\mathrm{ReLU}^2$ is an activation function that can be recognized as a simplified Softmax function in vanilla Transformers~\cite{so2021searching}. The output of Seq2map is the gated representation of $\boldsymbol{\widebar{Z}}$:
\begin{equation}
    \boldsymbol{\widehat{Z}} = \boldsymbol{\widebar{Z}} \; \odot \; \sigma(\boldsymbol{\widebar{Z}}),
\end{equation}
where $\sigma(\cdot)$ is the $\mathrm{Sigmoid}$ function that performs as a gate.

\section{Experiments}

We conduct experiments to compare our proposed RFold with state-of-the-art and commonly used approaches. Multiple experimental settings are taken into account, including standard structure prediction, generalization evaluation, large-scale benchmark evaluation, cross-family evaluation, pseudoknot prediction and inference time comparison. Detailed experimental setups can be found in the Appendix~\ref{sec:exp_details}.

\subsection{Standard RNA Secondary Structure Prediction}

Following~\citep{e2efold}, we split the RNAStralign dataset into train, validation, and test sets by stratified sampling. We report the results in Table~\ref{tab:rnastralign_test}. Energy-based methods achieve relatively weak F1 scores ranging from 0.420 to 0.633. Learning-based folding algorithms like E2Efold and UFold significantly improve performance by large margins, while RFold obtains even better performance among all the metrics. Moreover, RFold obtains about 8\% higher precision than the state-of-the-art method. This suggests that our optimization strategy is strict to satisfy all the hard constraints for predicting valid structures. 


\begin{table}[ht]
\centering
\caption{Results on RNAStralign test set. Results in bold and underlined are the top-1 and top-2 performances, respectively.}
\setlength{\tabcolsep}{3.5mm}{
\begin{tabular}{cccc}
\toprule
Method       & Precision & Recall & F1    \\
\midrule
Mfold        & 0.450     & 0.398  & 0.420 \\
RNAfold      & 0.516     & 0.568  & 0.540 \\
RNAstructure & 0.537     & 0.568  & 0.550 \\
CONTRAfold   & 0.608     & 0.663  & 0.633 \\
LinearFold   & 0.620     & 0.606  & 0.609 \\
CDPfold      & 0.633     & 0.597  & 0.614 \\
E2Efold      & 0.866     & 0.788  & 0.821 \\
UFold        & \underline{0.905} & \underline{0.927} & \underline{0.915} \\
\hline
\rowcolor{gray!10}
RFold        & \textbf{0.981} & \textbf{0.973} & \textbf{0.977} \\
\bottomrule
\end{tabular}}
\label{tab:rnastralign_test}
\end{table}

\subsection{Generalization Evaluation}

To verify the generalization ability of our proposed RFold, we directly evaluate the performance on another benchmark dataset ArchiveII using the pre-trained model on the RNAStralign training dataset. Following~\cite{e2efold}, we exclude RNA sequences in ArchiveII that have overlapping RNA types with the RNAStralign dataset for a fair comparison. The results are reported in Table~\ref{tab:archiveii}.

It can be seen that traditional methods achieve F1 scores in the range of 0.545 to 0.842. A recent learning-based method, MXfold2, obtains an F1 score of 0.768, which is even lower than some energy-based methods. Another state-of-the-art learning-based method improves the performance to the F1 score of 0.905. RFold further improves the F1 score to 0.921, even higher than UFold. It is worth noting that RFold has a relatively lower result in the recall metric and a significantly higher result in the precision metric. The reason for this phenomenon may be the strict constraints of RFold. While none of the current learning-based methods can satisfy all the constraints we introduced in Sec.~\ref{sec:constraints}, the predictions of RFold are guaranteed to be valid. Thus, RFold may cover fewer pair-wise interactions, leading to a lower recall metric. However, the highest F1 score still suggests the great generalization ability of RFold.

\begin{table}[ht]
\vspace{-4mm}
\centering
\caption{Results on ArchiveII dataset. Results in bold and underlined are the top-1 and top-2 performances, respectively.}
\setlength{\tabcolsep}{3.5mm}{
\begin{tabular}{cccc}
\toprule
Method       & Precision & Recall & F1    \\
\midrule
Mfold        & 0.668 & 0.590 & 0.621 \\
CDPfold      & 0.557 & 0.535 & 0.545 \\
RNAfold      & 0.663 & 0.613 & 0.631 \\
RNAstructure & 0.664 & 0.606 & 0.628 \\
CONTRAfold   & 0.696 & 0.651 & 0.665 \\
LinearFold   & 0.724 & 0.605 & 0.647 \\
RNAsoft      & 0.665 & 0.594 & 0.622 \\
Eternafold   & 0.667 & 0.622 & 0.636 \\
E2Efold      & 0.734 & 0.660 & 0.686 \\
SPOT-RNA     & 0.743 & 0.726 & 0.711 \\
MXfold2      & 0.788 & 0.760 & 0.768 \\
Contextfold  & 0.873 & 0.821 & 0.842 \\
RTfold       & \underline{0.891} & 0.789 & 0.814 \\
UFold        & 0.887 & \textbf{0.928} & \underline{0.905} \\
\hline
\rowcolor{gray!10}
RFold        & \textbf{0.938} & \underline{0.910} & \textbf{0.921} \\
\bottomrule
\end{tabular}}
\label{tab:archiveii}
\vspace{-4mm}
\end{table}

\subsection{Large-scale Benchmark Evaluation}

The bpRNA dataset is a large-scale benchmark, comprises fixed training (TR0), evaluation (VL0), and testing (TS0) sets. Following previous works~\cite{spot-rna,sato2021rna,ufold}, we train the model in bpRNA-TR0 and evaluate the performance on bpRNA-TS0 by using the best model learned from bpRNA-VL0. The detailed results can be found in Table~\ref{tab:bprna}.

\begin{table}[ht]
\vspace{-5mm}
\centering
\caption{Results on bpRNA-TS0 set.}
\setlength{\tabcolsep}{3.5mm}{
\begin{tabular}{cccc}
\toprule
Method       & Precision & Recall & F1    \\
\midrule
E2Efold      & 0.140 & 0.129  & 0.130 \\
RNAstructure & 0.494 & 0.622  & 0.533 \\
RNAsoft      & 0.497 & 0.626  & 0.535 \\
RNAfold      & 0.494 & 0.631  & 0.536 \\
Mfold        & 0.501 & 0.627  & 0.538 \\
Contextfold  & 0.529 & 0.607  & 0.546 \\
LinearFold   & 0.561 & 0.581  & 0.550 \\
MXfold2      & 0.519 & 0.646  & 0.558 \\
Externafold  & 0.516 & \underline{0.666}  & 0.563 \\
CONTRAfold   & 0.528 & 0.655  & 0.567 \\
SPOT-RNA     & \underline{0.594} & 0.693  & \underline{0.619} \\
UFold  & 0.521 & 0.588 & 0.553 \\
\hline
\rowcolor{gray!10}
RFold  & \textbf{0.692} & 0.635 & \textbf{0.644} \\
\bottomrule    
\end{tabular}}
\vspace{-4mm}
\label{tab:bprna}
\end{table}

RFold outperforms the prior state-of-the-art method, SPOT-RNA, by a notable 4.0\% in terms of the F1 score. This improvement in the F1 score can be attributed to the consistently superior performance of RFold in the precision metric when compared to baseline models. However, it is important to note that the recall metric remains constrained, likely due to stringent constraints applied during prediction.

\begin{table}[h]
\vspace{-5mm}
\centering
\caption{Results on long-range bpRNA-TS0 set. Results in bold and underlined are the top-1 and top-2 performances, respectively.}
\setlength{\tabcolsep}{3.5mm}{
\begin{tabular}{cccc}
\toprule
Method       & Precision & Recall & F1    \\
\midrule
Mfold        & 0.315     & 0.450  & 0.356 \\
RNAfold      & 0.304     & 0.448  & 0.350 \\
RNAstructure & 0.299     & 0.428  & 0.339 \\
CONTRAfold   & 0.306     & 0.439  & 0.349 \\
LinearFold   & 0.281     & 0.355  & 0.305 \\
RNAsoft      & 0.310     & 0.448  & 0.353 \\
Externafold  & 0.308     & 0.458  & 0.355 \\
SPOT-RNA     & 0.361     & 0.492  & 0.403 \\
MXfold2      & 0.318     & 0.450  & 0.360 \\
Contextfold  & 0.332     & 0.432  & 0.363 \\
UFold        & \underline{0.543} & \underline{0.631}  & \underline{0.584} \\
\hline
\rowcolor{gray!10}
RFold        & \textbf{0.803} & \textbf{0.765} & \textbf{0.701} \\
\bottomrule
\end{tabular}}
\label{tab:bprna_longrange}
\vspace{-2mm}
\end{table}

Following~\cite{ufold}, we conduct an experiment on long-range interactions. Given a sequence of length $L$, the long-range base pairing is defined as the paired and unpaired bases with intervals longer than $L/2$. As shown in Table~\ref{tab:bprna_longrange}, RFold performs unexpectedly well on these long-range base pairing predictions and improves UFold in all metrics by large margins, demonstrating its strong predictive ability.

\vspace{-3mm}
\subsection{Cross-family Evaluation}

\begin{table}[ht]
\vspace{-5mm}
\centering
\caption{Results on bpRNA-new. Results in bold and underlined are the top-1 and top-2 performances, respectively.}
\setlength{\tabcolsep}{3.5mm}{
\begin{tabular}{cccc}
\toprule
Method       & Precision & Recall & F1    \\
\midrule
E2Efold      & 0.047 & 0.031  & 0.036 \\ 
SPOT-RNA     & \textbf{0.635} & 0.641  & 0.620 \\
Contrafold   & \underline{0.620} & \underline{0.736}  & \textbf{0.661} \\
UFold  & 0.500 & 0.736 & 0.583 \\
UFold + aug & 0.570 & \textbf{0.742} & 0.636 \\
\hline
\rowcolor{gray!10}
RFold  & 0.614 & 0.619 & 0.616 \\
\rowcolor{gray!10}
RFold + aug  & 0.618 & 0.687 & \underline{0.651} \\
\bottomrule    
\end{tabular}}
\label{tab:bprna_new}
\vspace{-3mm}
\end{table}
    
The bpRNA-new dataset is a cross-family benchmark dataset comprising 1,500 RNA families, presenting a significant challenge for pure deep learning approaches. UFold, for instance, relies on the thermodynamic method Contrafold for data augmentation to achieve satisfactory results. As shown in Table~\ref{tab:bprna_new}, the standard UFold achieves an F1 score of 0.583, while RFold reaches 0.616. When the same data augmentation technique is applied, UFold's performance increases to 0.636, whereas RFold yields a score of 0.651. This places RFold second only to the thermodynamics-based method, Contrafold, in terms of F1 score.

\subsection{Predict with Pseudoknots}

Following E2Efold~\cite{e2efold}, we consider a sequence to be a true positive if it is correctly identified as containing a pseudoknot. For this analysis, we extracted all sequences featuring pseudoknots from the RNAStralign test dataset and assessed their predictive accuracy. The results of this analysis are summarized in the following table:
\begin{table}[h]
\vspace{-2mm}
\centering
\caption{Results on RNA structures with pseudoknots.}
\label{table:comparison}
\setlength{\tabcolsep}{3mm}{
\begin{tabular}{lccc}
\toprule
Method & Precision & Recall & F1 Score \\
\hline
RNAstructure & 0.778 & 0.761 & 0.769 \\
SPOT-RNA & 0.677 & 0.978 & 0.800 \\
E2Efold  & 0.844 & 0.990 & 0.911 \\
UFold  & 0.962 & 0.990 & 0.976 \\
\hline
\rowcolor{gray!10}
RFold & \textbf{0.971} & \textbf{0.993} & \textbf{0.982} \\
\bottomrule
\end{tabular}}
\vspace{-2mm}
\end{table}
   
RFold demonstrates superior performance compared to its counterparts across all evaluated metrics, i.e., precision, recall, and F1 score. This consistent outperformance across multiple dimensions of accuracy underscores the efficacy and robustness of the RFold approach in predicting RNA structures with pseudoknots.

\subsection{Inference Time Comparison}

We compared the running time of various methods for predicting RNA secondary structures using the RNAStralign testing set with the same experimental setting and the hardware environment as in~\citep{ufold}. The results are presented in Table~\ref{tab:inference_time}, which shows the average inference time per sequence. 
The fastest energy-based method, LinearFold, takes about 0.43s for each sequence. The learning-based baseline, UFold, takes about 0.16s. RFold has the highest inference speed, costing only about 0.02s per sequence. In particular, RFold is about eight times faster than UFold and sixteen times faster than MXfold2.

\begin{table}[ht]
\vspace{-3mm}
\centering
\caption{Inference time on the RNAStralign test set. Results in bold and underlined are the top-1 and top-2 performances, respectively.}
\setlength{\tabcolsep}{6mm}{
\begin{tabular}{cc}
\toprule
Method               & Time          \\
\midrule
CDPfold (Tensorflow) & 300.11 s      \\
RNAstructure (C)     & 142.02 s      \\
CONTRAfold (C++)     & 30.58 s       \\
Mfold (C)            & 7.65 s        \\
Eternafold (C++)     & 6.42 s        \\
RNAsoft (C++)        & 4.58 s        \\
RNAfold (C)          & 0.55 s        \\
LinearFold (C++)     & 0.43 s        \\
SPOT-RNA(Pytorch)    & 77.80 s (GPU) \\
E2Efold (Pytorch)    & 0.40 s (GPU)  \\
MXfold2 (Pytorch)    & 0.31 s (GPU)  \\
UFold (Pytorch)      & \underline{0.16 s} (GPU)  \\
\hline
\rowcolor{gray!10}
RFold (Pytorch)      & \textbf{0.02 s} (GPU) \\
\bottomrule
\end{tabular}}
\label{tab:inference_time}
\vspace{-2mm}
\end{table}

\subsection{Ablation Study}

\paragraph{Bi-dimensional Optimization} 

To validate the effectiveness of our proposed bi-dimensional optimization strategy, we conduct an experiment that replaces them with other optimization methods. The results are summarized in Table~\ref{tab:ablation_post}, where RFold-E and RFold-S denote our model with the optimization strategies of E2Efold and SPOT-RNA, respectively. While precision, recall, and F1 score are evaluated at base-level, we report the validity which is a sample-level metric evaluating whether the predicted structure satisfies all the constraints. It can be seen that though RFold-E has comparable performance in the first three metrics with ours, many of its predicted structures are invalid. The optimization strategy of SPOT-RNA has incorporated no constraint that results in its low validity. Moreover, its strategy seems to not fit our model well, which may be caused by the simplicity of our proposed RFold model.

\begin{table}[ht]
\centering
\caption{Ablation study on different optimization strategies (RNAStralign testing set).}
\setlength{\tabcolsep}{2.5mm}{
\begin{tabular}{ccccc}
\toprule
Method  & Precision & Recall & F1    & Validity \\
\midrule
\rowcolor{gray!10}
RFold   & \textbf{0.981} & \textbf{0.973} & \textbf{0.977} & \textbf{100.00}\% \\
\hline
RFold-E & 0.888 & 0.906 & 0.896 & 50.31\%  \\
RFold-S & 0.223 & 0.988 & 0.353 & 0.00\%   \\
\bottomrule
\end{tabular}}
\label{tab:ablation_post}
\end{table}

\paragraph{Seq2map Attention} We also conduct an experiment to evaluate the proposed Seq2map attention. We replace the Seq2map attention with the hand-crafted features from UFold and the outer concatenation from SPOT-RNA, which are denoted as RFold-U and RFold-SS, respectively. In addition to performance metrics, we also report the average inference time for each RNA sequence to evaluate the model complexity. We summarize the result in Table~\ref{tab:ablation_pre}. It can be seen that RFold-U takes much more inference time than our RFold and RFold-SS due to the heavy computational cost when loading and learning from hand-crafted features. Moreover, it is surprising to find that RFold-SS has a little better performance than RFold-U, with the least inference time for its simple outer concatenation operation. However, neither RFold-U nor RFold-SS can provide informative representations like our proposed Seq2map attention. With comparable inference time with the simplest RFold-SS, our RFold outperforms baselines by large margins.

\begin{table}[ht]
\centering
\caption{Ablation study on different pre-processing strategies (RNAStralign testing set).}
\setlength{\tabcolsep}{2.5mm}{
\begin{tabular}{ccccc}
\toprule
Method  & Precision & Recall & F1 & Time \\
\midrule
\rowcolor{gray!10}
RFold   & \textbf{0.981} & \textbf{0.973} & \textbf{0.977} & 0.0167 \\
\hline
RFold-U  & 0.875 & 0.941 & 0.906 & 0.0507 \\
RFold-SS & 0.886 & 0.945 & 0.913 &  \textbf{0.0158}  \\
\bottomrule
\end{tabular}}
\label{tab:ablation_pre}
\end{table}

\paragraph{Row-wise and Column-wise Componenets} We conducted comprehensive ablation studies on the row-wise and column-wise components of our proposed model, RFold, by modifying the inference mechanism using pre-trained checkpoints. These studies were meticulously designed to isolate and understand the individual contributions of these components to our model's performance in RNA secondary structure prediction. The results, presented across three datasets—RNAStralign (Table~\ref{tab:row_col_rnastralign}), ArchiveII (Table~\ref{tab:row_col_archiveii}), and bpRNA-TS0 (Table~\ref{tab:row_col_bprna})—highlight two key findings: (i) Removing both the row-wise and column-wise components leads to a substantial drop in the model's performance, underscoring their pivotal role within our model's architecture. This dramatic reduction in effectiveness clearly demonstrates that both components are integral to achieving high accuracy. The significant decline in performance when these components are omitted highlights their essential function in capturing the complex dependencies within RNA sequences. (ii) The performance metrics when isolating either the row-wise or column-wise components are remarkably similar across all datasets. This uniformity suggests that the training process, which incorporates row-wise and column-wise softmax functions, likely yields symmetric outputs. Consequently, this symmetry implies that each component contributes in an almost equal measure to the model's overall predictive capacity.

\begin{table}[ht]
\centering
\vspace{-4mm}
\caption{Ablation study on row-wise and column-wise components (RNAStralign testing set).}
\setlength{\tabcolsep}{1.8mm}{
\begin{tabular}{ccccc}
\toprule
Method  & Precision & Recall & F1    & Validity \\
\midrule
\rowcolor{gray!10}
RFold   & 0.981 & 0.973 & 0.977 & 100.00\% \\
\hline
RFold w/o C & 0.972 & 0.975 & 0.973 & 75.99\%  \\
RFold w/o R & 0.972 & 0.975 & 0.973 & 75.99\%   \\
RFold w/o R,C & 0.016 & 0.031 & 0.995 & 0.00\%   \\
\bottomrule
\end{tabular}}
\label{tab:row_col_rnastralign}
\end{table}

\begin{table}[ht]
\centering
\vspace{-4mm}
\caption{Ablation study on row-wise and column-wise components (ArchiveII).}
\setlength{\tabcolsep}{1.8mm}{
\begin{tabular}{ccccc}
\toprule
Method  & Precision & Recall & F1    & Validity \\
\midrule
\rowcolor{gray!10}
RFold   & 0.938 & 0.910 & 0.921 & 100.00\% \\
\hline
RFold w/o C & 0.919 & 0.914 & 0.914 & 49.14\%  \\
RFold w/o R & 0.919 & 0.914 & 0.914 & 49.14\%   \\
RFold w/o R,C & 0.013 & 0.997 & 0.025 & 0.00\%   \\
\bottomrule
\end{tabular}}
\label{tab:row_col_archiveii}
\end{table}

\begin{table}[ht]
\centering
\vspace{-4mm}
\caption{Ablation study on row-wise and column-wise components (bpRNA-TS0).}
\setlength{\tabcolsep}{1.8mm}{
\begin{tabular}{ccccc}
\toprule
Method  & Precision & Recall & F1    & Validity \\
\midrule
\rowcolor{gray!10}
RFold   & 0.693 & 0.635 & 0.644 & 100.00\% \\
\hline
RFold w/o C & 0.652 & 0.651 & 0.637 & 12.97\%  \\
RFold w/o R & 0.652 & 0.651 & 0.637 & 12.97\%   \\
RFold w/o R,C & 0.021 & 0.995 & 0.040 & 0.00\%   \\
\bottomrule
\end{tabular}}
\label{tab:row_col_bprna}
\end{table}

\subsection{Visualization}

\begin{figure}[h]
    \centering
    \includegraphics[width=0.48\textwidth]{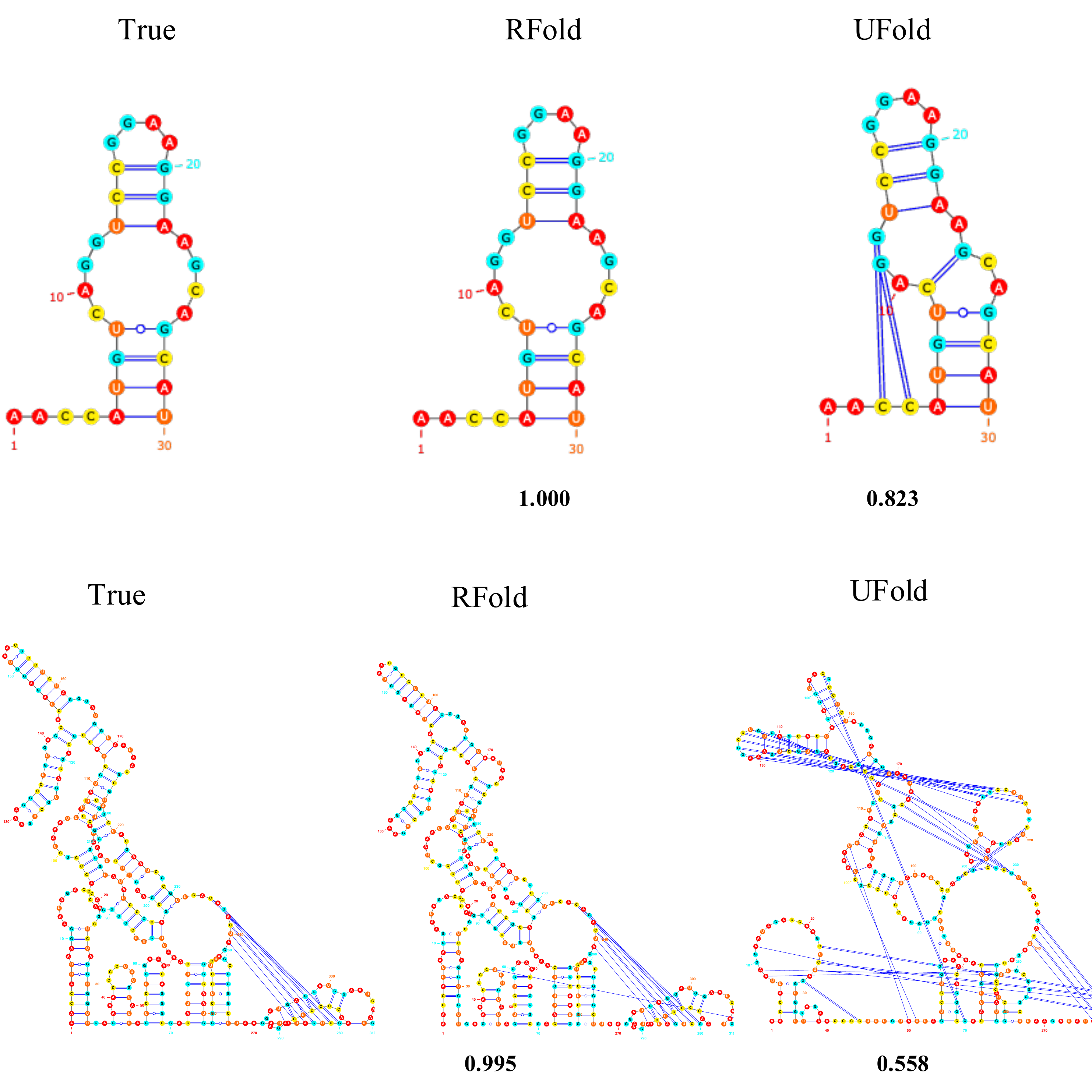}
    \vspace{-4mm}
    \caption{Visualization of the true and predicted structures.}
    \vspace{-4mm}
    \label{fig:visualization}
\end{figure}

We visualize two examples predicted by RFold and UFold in Figure~\ref{fig:visualization}. The corresponding F1 scores are denoted at the bottom right of each plot. The first row of secondary structures is a simple example of a nested structure. It can be seen that UFold may fail in such a case. The second row of secondary structures is much more difficult that contains over 300 bases of the non-nested structure. While UFold fails in such a complex case, RFold can predict the structure accurately. Due to the limited space, we provide more visualization comparisons in Appendix~\ref{sec:visual}.

\vspace{-2mm}

\section{Conclusion}

In this study, we reformulate RNA secondary structure prediction as a K-Rook problem, thus transforming the prediction process into probabilistic matching. Subsequently, we introduce RFold, an efficient learning-based model, which utilizes a bidimensional optimization strategy to decompose the probabilistic matching into row-wise and column-wise components, simplifying the solving process while guaranteeing the validity of the output. Comprehensive experiments demonstrate that RFold achieves competitive performance with faster inference speed. 

The limitations of RFold primarily revolve around its stringent constraints. This strictness in constraints implies that RFold is cautious in predicting interactions, leading to higher precision but possibly at the cost of missing some true interactions. Though we have provided a naive solution in Appendix~\ref{sec:abnormal_samples}, it needs further studies to obtain a better strategy that leads to more balanced precision-recall trade-offs and more comprehensive structural predictions.

\section*{Acknowledgements}

This work was supported by National Science and Technology Major Project (No. 2022ZD0115101), National Natural Science Foundation of China Project (No. U21A20427), Project (No. WU2022A009) from the Center of Synthetic Biology and Integrated Bioengineering of Westlake University and Integrated Bioengineering of Westlake University and Project (No. WU2023C019) from the Westlake University Industries of the Future Research Funding.

\section*{Impact Statement}

RFold is the first learning-based method that guarantees the validity of predicted RNA secondary structures. Its capability to ensure accurate predictions. It can be a valuable tool for biologists to study the structure and function of RNA molecules. Additionally, RFold stands out for its speed, significantly surpassing previous methods, marking it as a promising avenue for future developments in this field. There are many potential societal consequences of our work, none of which we feel must be specifically highlighted here.

\bibliography{ref}
\bibliographystyle{icml2024}


\appendix
\clearpage

\section{Comparison of mainstream RNA secondary structure prediction methods}
\label{sec:comparison_methods}

\begin{table*}[ht]
\small
\centering
\caption{Comparison between RNA secondary structure prediction methods and RFold.}
\setlength{\tabcolsep}{2mm}{
\begin{tabular}{ccccc}
\toprule
Method                             & SPOT-RNA        & E2Efold                   & UFold                     & RFold                     \\
\midrule
pre-processing            & pairwise concat & pairwise concat           & hand-crafted              & seq2map attention         \\
optimization approach           & $\times$ & unrolled algorithm & unrolled algorithm & bi-dimensional optimization \\
\midrule
constraint (a)                     & $\times$        & \checkmark & \checkmark & \checkmark \\
constraint (b)                     & $\times$        & \checkmark & \checkmark & \checkmark \\
constraint (c)                     & $\times$        & $\times$                  & $\times$                  & \checkmark \\
\midrule
F1 score             & 0.711           & 0.821                     & 0.915                     & \textbf{0.977}                     \\
Inference time & 77.80 s         & 0.40 s                    & 0.16 s                    & \textbf{0.02} s   \\
\bottomrule                
\end{tabular}}
\label{tab:comparison_methods}
\end{table*}

We compare our proposed method RFold with several other leading RNA secondary structure prediction methods and summarize the results in Table~\ref{tab:comparison_methods}.
RFold satisfies all three constraints (a)-(c) for valid RNA secondary structures, while the other methods do not fully meet some of the constraints. RFold utilizes a sequence-to-map attention mechanism to capture long-range dependencies, whereas SPOT-RNA simply concatenates pairwise sequence information and E2Efold/UFold uses hand-crafted features.
In terms of prediction accuracy on the RNAStralign benchmark test set, RFold achieves the best F1 score of 0.977, outperforming SPOT-RNA, E2Efold and UFold by a large margin. Regarding the average inference time, RFold is much more efficient and requires only 0.02 seconds to fold the RNAStralign test sequences. 
In summary, RFold demonstrates superior performance over previous methods for RNA secondary structure prediction in both accuracy and speed. 

\section{Experimental Details}
\label{sec:exp_details}

\paragraph{Datasets} We use three benchmark datasets: (i) RNAStralign~\citep{tan2017turbofold}, one of the most comprehensive collections of RNA structures, is composed of 37,149 structures from 8 RNA types; (ii) ArchiveII~\citep{sloma2016exact}, a widely used benchmark dataset in classical RNA folding methods, containing 3,975 RNA structures from 10 RNA types; (iii) bpRNA~\citep{spot-rna}, is a large scale benchmark dataset, containing 102,318 structures from 2,588 RNA types. (iv) bpRNA-new~\citep{sato2021rna}, derived from Rfam 14.2~\citep{kalvari2021rfam}, containing sequences from 1500 new RNA families.

\paragraph{Baselines} We compare our proposed RFold with baselines including energy-based folding methods such as Mfold~\citep{zuker2003mfold}, RNAsoft~\citep{andronescu2003rnasoft}, RNAfold~\citep{lorenz2011viennarna}, RNAstructure~\citep{mathews2006prediction}, CONTRAfold~\citep{do2006contrafold}, Contextfold~\citep{zakov2011rich}, and LinearFold~\citep{huang2019linearfold}; learning-based folding methods such as SPOT-RNA~\citep{spot-rna}, Externafold~\citep{wayment2021rna}, E2Efold~\citep{e2efold}, MXfold2~\citep{sato2021rna}, and UFold~\citep{ufold}. 

\paragraph{Metrics} We evaluate the performance by precision, recall, and F1 score, which are defined as:
\begin{equation}
\small
\begin{aligned}  
\mathrm{Precision} &= \frac{\mathrm{TP}}{\mathrm{TP}+\mathrm{FP}}, \; \mathrm{Recall} = \frac{\mathrm{TP}}{\mathrm{TP}+\mathrm{FN}},\\
\mathrm{F1} &= 2 \; \frac{\mathrm{Precision} \cdot \mathrm{Recall}}{\mathrm{Precision} + \mathrm{Recall}},
\end{aligned}
\end{equation}
where $\mathrm{TP}, \mathrm{FP}$, and $\mathrm{FN}$ denote true positive, false positive and false negative, respectively.

\paragraph{Implementation details} Following the same experimental setting as~\citep{ufold}, we train the model for 100 epochs with the Adam optimizer. The learning rate is 0.001, and the batch size is 1 for sequences with different lengths.

\section{Discussion on Abnormal Samples}
\label{sec:abnormal_samples}

Although we have illustrated three hard constraints in~\ref{sec:constraints}, there exist some abnormal samples that do not satisfy these constraints in practice. We have analyzed the datasets used in this paper and found that there are some abnormal samples in the testing set that do not meet these constraints. The ratio of valid samples in each dataset is summarized in the table below:

\begin{table}[ht]
\centering
\caption{The ratio of valid samples in the datasets.}
\setlength{\tabcolsep}{3mm}{
\begin{tabular}{cccc}
\toprule
Dataset  & RNAStralign & ArchiveII & bpRNA   \\
\midrule
Validity & 93.05\%     & 96.03\%   & 96.51\% \\
\bottomrule
\end{tabular}}
\end{table}

As shown in Table~\ref{tab:ablation_post}, RFold forces the validity to be 100.00\%, while other methods like E2Efold only achieve about 50.31\%. RFold is more accurate than other methods in reflecting the real situation.

Nevertheless, we provide a soft version of RFold to relax the strict constraints. A possible solution to relax the rigid procedure is to add a checking mechanism before the Argmax function in the inference. Specifically, if the confidence given by the Softmax is low, we do not perform Argmax and assign more base pairs. It can be implemented as the following pseudo-code:

\begin{lstlisting}[language=Python]
y_pred = row_col_softmax(y)
int_one = row_col_argmax(y_pred)

# get the confidence for each position
conf = y_pred * int_one
all_pos = conf > 0.0

# select reliable position
conf_pos = conf > thr1

# select unreliable position with the full row and column
uncf_pos = get_unreliable_pos(all_pos, conf_pos)

# assign "1" for the positions with the confidence higher than thr2
# note that thr2 < thr1
y_pred[uncf_pos] = (y_pred[uncf_pos] > thr2).float()
int_one[uncf_pos] = y_pred[uncf_pos]
\end{lstlisting}
\vspace{-2mm}

We conduct experiments to compare the soft-RFold and the original version of RFold in the RNAStralign dataset. The results are summarized in the Table~\ref{tab:soft_rfold_rnastralign}. It can be seen that soft-RFold improves the recall metric by a small margin. The minor improvement may be because the number of abnormal samples is small. We then select those samples that do not obey the three constraints to further analyze the performance. The total number of such samples is 179. It can be seen that soft-RFold can deal with abnormal samples well. The improvement of the recall metric is more obvious.

\begin{table}[ht]
  \begin{minipage}{.48\textwidth}
    \centering
    \caption{The results of soft-RFold and RFold on the RNAStralign.}
    \setlength{\tabcolsep}{2.5mm}{
    \begin{tabular}{cccc}
      \toprule
      Method  & Precision & Recall & F1   \\
      \midrule
      RFold   & 0.981     & 0.973  & 0.977 \\
      soft-RFold & 0.978  & 0.974  & 0.976 \\
      \bottomrule
      \end{tabular}}
    \label{tab:soft_rfold_rnastralign}
  \end{minipage}
  \quad \quad 
  \begin{minipage}{.48\textwidth}
    \centering
    \caption{The results of soft-RFold and RFold on the abnormal samples on the RNAStralign.}
    \setlength{\tabcolsep}{2.5mm}{
    \begin{tabular}{cccc}
      \toprule
      Method  & Precision & Recall & F1   \\
      \midrule
      RFold      & 0.956  & 0.860  & 0.905 \\
      soft-RFold & 0.949  & 0.889  & 0.918 \\
      \bottomrule
      \end{tabular}}
    \label{tab:soft_rfold_rnastralign_plus}
  \end{minipage}
\end{table}

\newpage

\section{Visualization}
\label{sec:visual}

\begin{figure*}[h]
    \centering
    \includegraphics[width=0.98\textwidth]{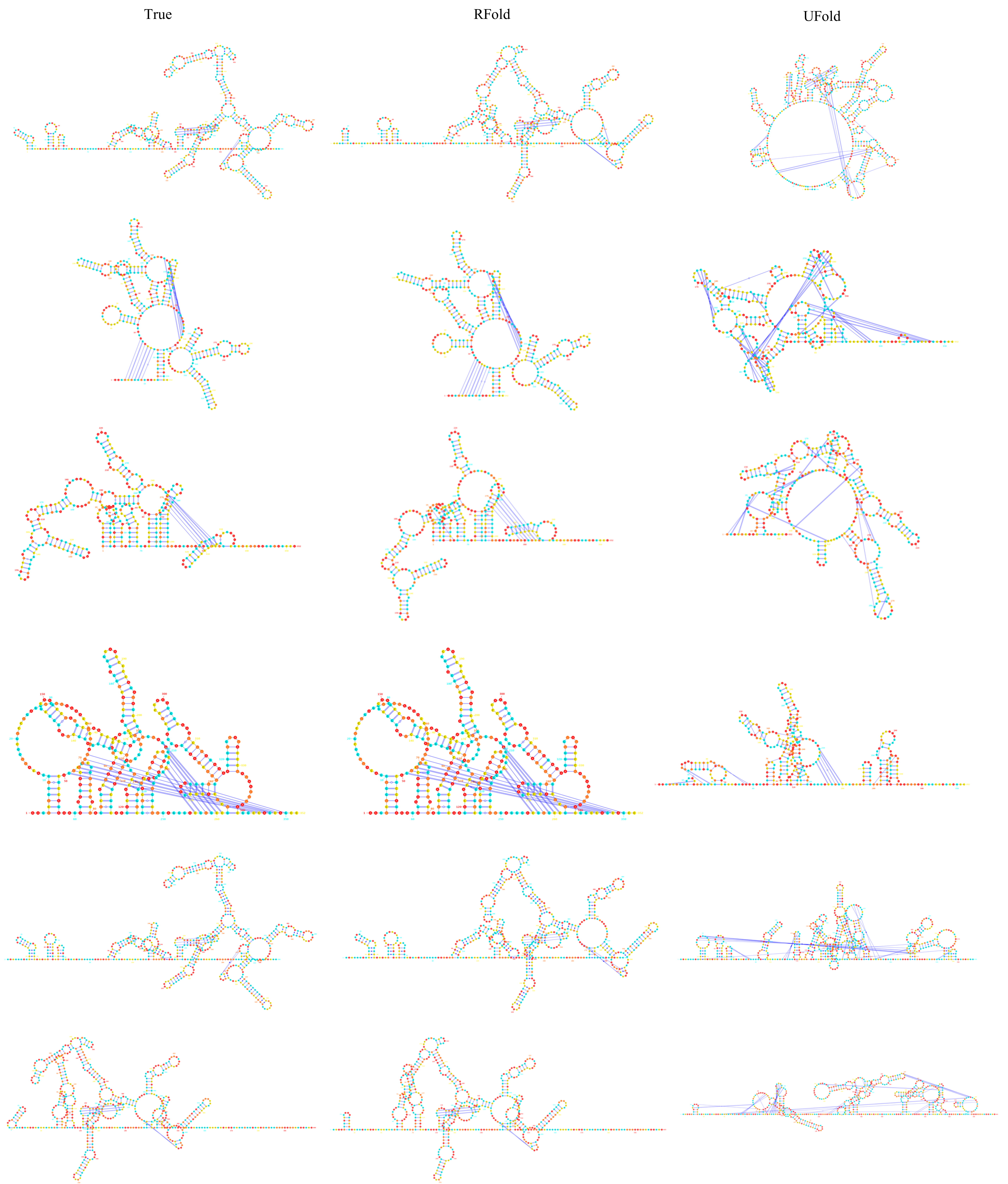}
    \caption{Visualization of the true and predicted structures.}
    \label{fig:visualization_appendix}
\end{figure*}

\end{document}